\documentclass[12pt,a4paper]{article}

\usepackage[]{graphicx}

\usepackage[latin1]{inputenc}
\usepackage{amsmath}
\usepackage{amsfonts}
\usepackage{amssymb}

\title{The $q$-exponential family in statistical physics}
\author{
Jan Naudts\\
\small Departement Natuurkunde, Universiteit Antwerpen,\\
\small Universiteitsplein 1, 2610 Antwerpen, Belgium\\
\small E-mail Jan.Naudts@ua.ac.be
}
\newcommand{\be}{\begin{eqnarray}}
\newcommand{\ee}{\end{eqnarray}}

\def\Tr{{\rm Tr\,}}
\def\url#1{{\tt \scriptsize #1}}

\begin{document}
\maketitle

\begin{abstract}
The notion of generalised exponential family is considered
in the restricted context of nonextensive statistical physics.
Examples are given of models belonging to this family.
In particular, the q-Gaussians are discussed and it is shown
that the configurational probability distributions of the
microcanonical ensemble belong to the q-exponential family.
\end{abstract}

\noindent
KEYWORDS
\emph{
Q-exponential family, deformed logarithm, q-Gaussian,
kappa-distribution, microcanonical ensemble, variational principle,
maximum entropy principle, information geometry, Fisher information,
statistical manifold.
}

\section{Introduction}

In statistics, a model is a probability distribution which depends
on a number of parameters. With this definition, statistical physics
is plenty of models. Typical parameters are the total energy $U$
in the microcanonical ensemble, or the inverse temperature $\beta$
in the canonical ensemble. This parameter dependence is important
to understand why certain models belong to the exponential family
and others do not. In particular, all models described by
a Boltzmann-Gibbs distribution
\be
f_\beta(x)=\frac {c(x)}{Z(\beta)}\exp(-\beta H(x))
=c(x)\exp(-\ln Z(\beta)-\beta H(x)),
\label {intro:BG}
\ee
where $H(x)$ is the Hamiltonian of the system and $Z(\beta)$ is the
normalisation, belong to the
exponential family because they have the right dependence on
the inverse temperature $\beta$.

Recently, the notion of the exponential family has been
generalised by the present author in a series of papers
\cite {NJ04,NJ05,NJ06,NJ08}.
The same definition of generalised exponential family
has also been introduced in the mathematics literature
\cite {NJ04,GD04,SV07,NJ08}.
This class of models was also derived using the
maximum entropy principle in \cite {AS03,HT07}.

Many but not all of the models of non-extensive statistical physics \cite {TC88,TC04}
belong to the generalised exponential family.
They are obtained by replacing in (\ref {intro:BG}) the exponential function
by a $q$-deformed exponential function \cite {TC94,NJ02}
--- see the next Section. An important question is then
whether in the modification the normalisation
should stand in front of the deformed exponential function,
or whether it should be included as $\ln Z(\beta)$ inside.
From the general formalism mentioned above it follows that the
latter is the right way to go. It is the intention of the
present paper to give examples and to show how the generalised formalism looks like
when restricted to the context of non-extensive statistical physics.

The next Sections recall the definition of deformed logarithmic and exponential functions and
introduce the notion of the $q$-exponential family. In Section 4, a number of physically
relevant examples are discussed. Sections 5 to 8 give the proof of the variational
principle. Sections 9 to 13 discuss the geometrical structure behind the $q$-exponential family.
In Section 14 some final remarks are made. The short appendix contains a table with
often used formulas.

\section{Deformed logarithmic and exponential functions}

The $q$-deformed logarithm was introduced in \cite {TC94}.
It is defined by
\be
\ln_q(u)=\frac 1{1-q}\left(u^{1-q}-1\right),
\qquad u>0.
\ee
Its first derivative is
\be
\frac {{\rm d}\,}{{\rm d}u}\ln_q(u)=\frac 1{u^q}
\ee
This derivative is positive for any value of $q$. Hence, the deformed logarithm is always a strictly increasing function
--- this is important in the sequel.
In the limit $q=1$ the deformed logarithm reduces to the natural logarithm $\ln u$.

The inverse function is the deformed exponential function
\be
\exp_q(u)=\left[1+(1-q)u\right]_+^{1/(1-q)}.
\ee
The notation $[u]_+=\max\{0,u\}$ is used. One has $0\le\exp_q(u)\le+\infty$ for all $u$.
For $q\not=1$ the range of $\ln_q(u)$ is not the full line. By putting $\exp_q(u)=0$ when
$u$ is below the range of $\ln_q(u)$, and equal to $+\infty$ when it is above,
$\exp_q(u)$ is an increasing function of $u$, defined for all values of $u$.

\section{The $q$-exponential family}

Some interesting models of statistical physics can be written into the following form
\be
f_\beta(x)
&=&c(x)\exp_q\left(-\alpha(\beta)-\beta H(x)\right).
\label {qexp:def}
\ee
If the $q$-exponential in the r.h.s.~diverges then $f_\beta(x)=0$ is assumed.
The function $H(x)$ is the Hamiltonian.
The parameter $\beta$ is usually the inverse temperature.
The normalisation $\alpha(\beta)$ is written inside the $q$-exponential.
The function $c(x)$ is the prior distribution.
It is a reference measure and must not depend on the
parameter $\beta$.

If a model is of the above form then it is said to belong to the $q$-exponential family.
In the limit $q=1$ these are the models of the standard exponential family.
In that case the expression (\ref {qexp:def}) reduces to
\be
f_\beta(x)=c(x)\exp\left(-\alpha(\beta)-\beta H(x)\right),
\ee
which is known as the Boltzmann-Gibbs distribution.

The convention that $f_\beta(x)=0$ when the r.h.s.~of (\ref {qexp:def})
diverges may seem weird. However, one can argue \cite {NJ08}
that this is the right thing to do. Also, the example of the harmonic
oscillator, given below, will clarify this point. A reformulation of (\ref {qexp:def})
is therefore (See Theorem 2 of \cite {NJ08}) that either $f_\beta(x)=0$ or
\be
\ln_q\left(\frac {f_\beta(x)}{c(x)}\right)=-\alpha(\beta)-\beta H(x).
\label {qexpfam:lnexpr}
\ee

The $q$-exponential family is a special case of the
generalised exponential family introduced in \cite {NJ04,GD04,NJ08}.
Models belonging to such a family share a number of interesting properties.
In particular, they all fit into the thermodynamic formalism.
As a consequence, the probability density $f_\beta(x)$ may be considered to
be the equilibrium distribution of the model at the given value of the parameter $\beta$.

\section{Examples}

\subsection{The $q$-Gaussian distribution ($q<3$)}

\begin{figure}
\begin {center}
\includegraphics[width=6cm]{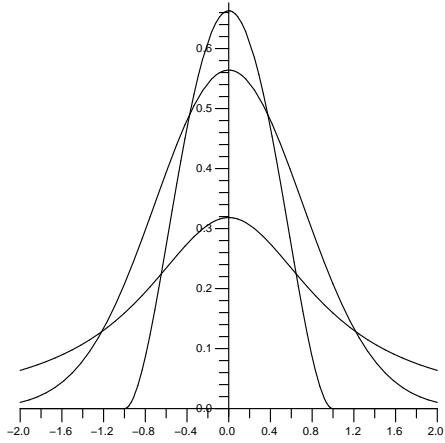}
\end {center}
\caption{$q$-Gaussians for $\sigma=1$ and $q=\frac 12,q=1,q=2$.}
\label {fig:qgauss}
\end{figure}

Many of the distributions encountered in the literature on nonextensive statistics
(see for instance \cite {TC04})
can be brought into the form  (\ref {qexp:def}).
A prominent model encountered in this context is the $q$-Gaussian distribution
(see for instance \cite {MTG06,HS07,TH08} and \cite {UT07,VP07,VP07b})
\be
f(x)=\frac {1}{c_q \sigma}\exp_q(- x^2/\sigma^2),
\label {eg:qgauss}
\ee
with
\be
c_q&=&\int_{-\infty}^\infty{\rm d}x\,\exp_q(- x^2)\crcr
&=& \sqrt{\frac {\pi}{q-1}}\frac{\Gamma\left(-\frac 12+\frac 1{q-1}\right)}{\Gamma\left(\frac 1{q-1}\right)}
\quad \mbox { if }\quad 1<q<3,\crcr
&=& \sqrt{\frac {\pi}{1-q}}\frac{\Gamma\left(1+\frac 1{1-q}\right)}{\Gamma\left(\frac 32+\frac 1{1-q}\right)}
\quad \mbox { if } \quad q<1.
\ee
It can be brought into the form (\ref {qexp:def}) with
$c(x)=1/c_q$, $H(x)=x^2$, $\beta=\sigma^{3-q}$, and
\be
\alpha(\beta)=\frac {\sigma^{q-1}-1}{q-1}=\ln_{2-q}(\sigma).
\ee

Take for instance $q=1/2$. Then (\ref{eg:qgauss}) becomes
\be
f(x)&=&\frac {15\sqrt 2}{32\sigma}\left[1-\frac {x^2}{\sigma^2}\right]_+^2.
\ee
Note that this distribution vanishes outside the interval $[-\sigma,\sigma]$.
The $q=1$-case reproduces the conventional Gauss distribution.
For $q=2$ one obtains
\be
f(x)=\frac 1\pi \frac {\sigma}{x^2+\sigma^2}.
\label {cauchy}
\ee
This is known as the Cauchy distribution. The function (\ref {cauchy})
is also called a Lorentzian.
In the range $1\le q< 3$ the $q$-Gaussian is strictly positive on the whole
line. For $q<1$ it is zero outside an interval. For $q\ge 3$ the distribution
cannot be normalised because
\be
f(x)\sim\frac 1{|x|^{2/(q-1)}}
\mbox{ as }|x|\rightarrow\infty.
\ee

\subsection{Kappa-distributions $1<q<\frac 53$}

The following distribution is known in plasma physics as the kappa-distribution
--- see for instance \cite {MVMH95}
\be
f(v)=\frac 1{A(\kappa)v_0^3}
\frac {v^2}{\left(1+\frac 1{\kappa-a}\frac {v^2}{v_0^2}\right)^{1+\kappa}}.
\label {kappa:def}
\ee
This distribution is a modification of the Maxwell distribution, exhibiting a power law
decay like $f(v)\simeq v^{-2\kappa}$ for large $v$.

Expression (\ref {kappa:def}) can be written  in the form of a $q$-exponential with
$q=1+\frac 1{1+\kappa}$ and
\be
f(v)=\frac {v^2}{A(\kappa)v_0^3}
\exp_q\left(-\frac 1{2-q-(q-1)a}\frac {v^2}{v_0^2}\right).
\label {kappa:temp}
\ee
However, in order to be of the form (\ref {qexp:def}) ,
the prefactor of (\ref{kappa:temp}) should not depend on the parameter $v_0$.
Introduce an arbitrary constant $c>0$ with the dimensions of a velocity. Then one can write
\be
f(v)&=&\frac {4\pi v^2}{c^3}\exp_q\bigg(
-\ln_{2-q}\left[4\pi A(\kappa)\left(\frac {v_0}c\right)^3\right]\cr
& &
-\left[4\pi A(\kappa)\left(\frac {v_0}c\right)^3\right]^{q-1}
\frac 1{2-q-(q-1)a}\frac {v^2}{v_0^2}
\bigg).
\ee
This is now of the form (\ref {qexp:def})  with prior distribution
$c(v)=4\pi v^2/c^3$  and Hamiltonian $H(v)=\frac 12 m v^2$.
The inverse temperature $\beta$ is given by
\be
\beta=\left(4\pi A(\kappa)\left(\frac {v_0}c\right)^3\right)^{q-1}
\frac 2{2-q-(q-1)a}\frac 1{mv_0^2}.
\ee
In the $q=1$-limit one obtains the Maxwell distribution with $\beta=2/mv_0^2$,
as it should be. If $q>1$ then the inverse temperature $\beta$ depends on the
choice of the arbitrary constant $c$, while the distribution function does not
depend on $c$. This is rather disturbing since it means that a fit of (\ref {kappa:def})
to experimental data does not result in an absolute value for the inverse temperature
$\beta$.

The inequality $\kappa>\frac 12$ is required to make $f(v)$ normalisable.
This implies that $1<q<\frac 53$. From $\beta\sim v_0^{2-3(q-1)}$ then follows
that $\beta$ is a monotonically decreasing function of the average velocity $v_0$,
as it should be.

\subsection {Speed of the harmonic oscillator ($q=3$)}

\begin{figure}
\begin {center}
\includegraphics[width=8cm,height=6cm]{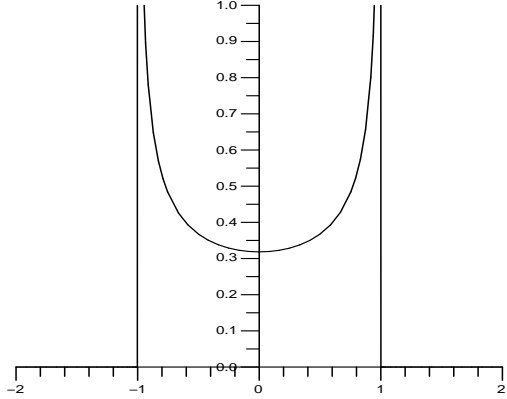}
\end {center}
\caption{Speed distribution for a harmonic oscillator with $v_0=1$.}
\label {fig:harmoscspeed}
\end{figure}

The distribution of velocities $v$ of a classical harmonic oscillator is given by
\be
f(v)=\frac 1\pi\frac 1{\sqrt{v_0^2-v^2}}.
\label {eg:ho}
\ee
It diverges when $|v|$ approaches its maximal value $v_0$ and vanishes for $|v|>v_0$.
See the Figure \ref {fig:harmoscspeed}. This distribution can be written into the form (\ref {qexp:def}) 
of a $q$-exponential family with $q=3$. To do so, let $x=v$ and
\be
c(x)&=&\frac {\sqrt 2}{\pi |v|},\cr
\beta&=&\frac 12mv_0^2,\cr
H(x)&=&\frac 1{\frac 12mv^2}\equiv\frac 1K,\cr
\alpha(\beta)&=&-\frac 32.
\ee
Remark that in this example the roles of the inverse temperature
and the Hamiltonian $H(x)$ are interchanged. The parameter $\beta$
in this model is the total energy $\frac 12mv_0^2$ of the harmonic
oscillator. The stochastic variable $H$, used to estimate the total energy,
is the inverse of the kinetic energy $K$.
Note however that the average of the latter diverges.

\subsection{Configurational density distribution ($q=-1$)}

\def\pp{{\bf p}}
\def\qq{{\bf q}}

The harmonic oscillator is a very special example because its density of states
\be
\rho(U)=\frac 1{2h}\int{\rm d}q\int{\rm d}p\,\delta(H(q,p)-U)
\ee
is constant and because it is quadratic in both the position and the momentum variables.
because of the latter property the role of the kinetic energy and the potential
energy can be interchanged. This is what is done below.

Consider a $d=3$-classical particle with mass $m$ in a potential $V(\qq)$.
The Hamiltonian is
\be
H(\qq,\pp)=\frac 1{2m}|\pp|^2+V(\qq).
\ee
The microcanonical probability distribution equals
\be
f_U(\qq,\pp)=\frac 1{\rho(U)}\delta(H(\qq,\pp)-U).
\ee
The configurational density distribution is obtained by integrating out the momenta
\be
f_U(\qq)
&=&\frac 1{\rho(U)}\frac 1{2h}\int{\rm d}\pp\,\delta(H(\qq,\pp)-U)\crcr
&=&\frac 1{\rho(U)}\frac 1{2h}\int_0^\infty 4\pi r^2{\rm d}r\,\delta\left(\frac 1{2m}r^2+V(\qq)-U\right)\crcr
&=&\frac 1{\rho(U)}\frac {2\sqrt 2 \pi m\sqrt m}{h}[U-V(\qq)]_+^{1/2}\crcr
&=&c\exp_{-1}\left(-\frac 12+\frac {U}{2\rho(U)^2}-\frac {V(\qq)}{2\rho(U)^2}\right),
\ee
with $c=2\sqrt 2 \pi m\sqrt m/h$.
Note that $f_U(\qq)$ is now in the form (\ref {qexp:def})
with $x=\qq$, $c(x)=c$, and
\be
\beta&=&\frac 1{2\rho(U)^2},\\
\alpha(\beta)&=&-\frac 12+\beta U(\beta),\\
H(x)&=&V(\qq).
\ee
Hence, the probability distribution of the position $q$ of the particle
belongs to the $q$-exponential family with $q=-1$.
The correct interpretation of this result is that the measured values of
\be
\langle V\rangle_\beta=\int{\rm d}\qq\,f_U(\qq)V(\qq)
\ee
can be used to estimate the parameter $\beta$. The latter determines the
original parameter of the microcanonical model, which is the total energy $U(\beta)$.
Indeed, assume that the density of states $\rho(U)$ is a strictly increasing function of $U$.
This is for instance the case when $V(\qq)\sim |\qq|^2$, which implies $\rho(U)\sim U^2$.
Then the function $\rho(U)$ can be inverted and the knowledge of $\beta$ uniquely determines
the total energy $U$.

\section{The variational principle}

An important argument justifying the statement that the model distributions (\ref {qexp:def}) 
exhibit statistical equilibrium is that they formally satisfy a maximum entropy principle.

It is known since long \cite {RD67} that the probability distributions of a model belonging to
the exponential family satisfy not only the maximum entropy principle, but also a stronger statement,
which is known in the mathematical physics literature as the variational principle. In physical
terms this principle states that the free energy is minimal in equilibrium.

The thermodynamic definition of free energy is $F=U-TS$, where $U$ is the average energy
$\langle H\rangle$, $S$ is the entropy, and $T$ is the temperature (the inverse of $\beta$ when
units are taken so that the Boltzmann constant equals 1).
It is slightly more convenient to maximise $\Phi=S-\beta U$ instead of minimising $F$.
This function $\Phi$ is known as Massieu's function.

In what follows it is shown that the model distributions (\ref {qexp:def}) satisfy
a generalised version of the variational principle. In the cases that the average energy
$U$ diverges the variational principle is satisfied only at the level of the microstates.

\section{Choice of the entropy function}

A general form of entropy function $I(f)$ is
\cite {AS03,NJ04,HT07,HT07b,NJ08}
\be
I(f)=-\int c(x){\rm d}x\,F\left(\frac{f(x)}{c(x)}\right),
\label {choice:genent}
\ee
with
\be
F(u)&=&\int_1^u{\rm d}v\,\Lambda(v)+A.
\ee
The function $\Lambda(v)$ should be a strictly increasing function
for  $I(f)$ to be an entropy function.
It may be interpreted as a deformed logarithm.
Hence, it is obvious to take $\Lambda(v)=\ln_q(v)$.
The result is then
\be
F_q(u)&=&\int_1^u{\rm d}v\ln_q(v)+A\cr
&=&\frac 1{1-q}\left(\frac 1{2-q}[u^{2-q}-1]-[u-1]\right)+A.
\label {varprin:Fdef}
\ee
The corresponding entropy function is denoted $I_q(f)$.

The constant $A$ in (\ref {varprin:Fdef}) is not yet determined.
Conventionally, it is chosen so that $F_q(0)=F_q(1)=0$.
This is only possible when $q=1$. Moreover, $F_q(0)$ diverges for $q>2$.
If $q<2$ then it is obvious to choose $A$ so that
\be
F_q(u)&=&\int_0^u{\rm d}v\ln_q(v)\cr
&=&\frac u{1-q}\left(\frac 1{2-q}u^{1-q}-1\right).
\label {varprin:Fdef2}
\ee
This choice of the function $F(u)$ in (\ref {choice:genent}) reproduces the
Tsallis entropy \cite {TC88} up to two modifications: a change of $q$ by
$2-q$ and the additional factor $1/(2-q)$ in front of $u^{1-q}$.

\section{Variational principle on the level of microstates}

\begin{figure}
\begin {center}
\includegraphics[width=6cm]{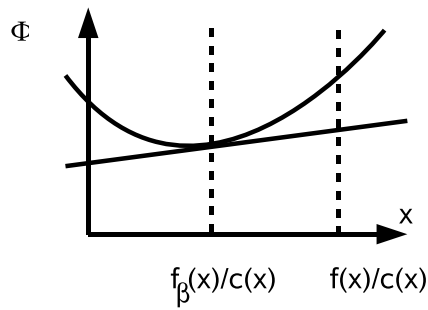}
\end {center}
\caption{Basic property of a convex function.}
\label {fig:convexity}
\end{figure}

Let be given a model with probability distributions of the form (\ref {qexp:def})
and fix one microstate $x$. Then each probability density $f(x)$ defines
a function $ M_{x,f}(\beta)$ by
\be
\beta\rightarrow M_{x,f}(\beta)\equiv
-c(x)F_q\left(\frac{f(x)}{c(x)}\right)
-[\alpha(\beta)+\beta H(x)]f(x).
\label {micro:max}
\ee
See the Figure \ref {fig:convexity}.
It is now easy to prove that
$M_{x,f}(\beta)\le M_{x,f_\beta}(\beta)$ for all $\beta$ for which $f_\beta(x)>0$.
In other words, the maximum in (\ref {micro:max}) is realised by
the equilibrium probability distributions of the model.

The proof goes as follows. The function $F_q(u)$ is convex. Hence, its value
at the point $f(x)/c(x)$ lies above the straight line which is tangent to
the function at the point $f_\beta(x)/c(x)$. See the Figure \ref {micro:max}.
In formulas this gives
\be
F_q\left(\frac{f(x)}{c(x)}\right)
\ge F_q\left(\frac{f_\beta(x)}{c(x)}\right)
+F_q'\left(\frac{f_\beta(x)}{c(x)}\right) \frac {f(x)-f_\beta(x)}{c(x)}.
\ee
Use now $F'_q(u)=\ln_q(u)$ in combination with (\ref {qexpfam:lnexpr}) to obtain
\be
F_q\left(\frac{f(x)}{c(x)}\right)
\ge F_q\left(\frac{f_\beta(x)}{c(x)}\right)
+[-\alpha(\beta)-\beta H(x)]\frac {f(x)-f_\beta(x)}{c(x)}.
\ee
This can be written as
\be
M_{x,f}(\beta)\le M_{x,f_\beta}(\beta)\equiv M_x(\beta).
\label {micro:ineq}
\ee

The inequality  (\ref {micro:max}) holds for all examples, even when the
average $\langle H\rangle_\beta$ diverges. Take for instance the
$q=3$-example of the speed of the harmonic oscillator. One finds
\be
M_{v,f}(\beta)&=&\left(1-\frac {2\beta}{mv^2}\right)f(v)
-\frac 1{\pi^2v^2 f(v)}+A-1.
\ee
This expression is maximal when $f_\beta(v)$ is given by (\ref {eg:ho}).

If $q=1$ then the equilibrium value $M_x(\beta)$ is identically zero. The variational principle then says that
for any probability density $f(x)$
\be
-f(x)\ln\frac {f(x)}{c(x)}+[-\alpha(\beta)-\beta H(x)]f(x)\le 0,
\ee
with equality when $f(x)=f_\beta(x)$.

\section{Proof of the variational principle}

It is now easy to prove the variational principle. Assume that $H(x)$ is bounded from below
and that the expectation value
\be
\langle H\rangle_\beta=\int{\rm d}x\,f_\beta(x)H(x)
\ee
converges. Then integration of (\ref {micro:max})  gives
\be
-\infty\le \int{\rm d}x\,M_{x,f}(\beta)
= I_q(f)-\alpha(\beta)-\beta \int{\rm d}x\,f(x)H(x).
\label {proof:mint}
\ee
The inequality (\ref {micro:ineq}) implies that (\ref {proof:mint})
is maximal when $f$ equals $f_\beta$.
Because $\alpha(\beta)$ does not depend on $f$
it may be subtracted. The statement then says that
\be
I_q(f)-\beta\int{\rm d}x\,f(x)H(x).
\label {proof:varprin}
\ee
is maximal when
$f$ equals $f_\beta$ as given by (\ref {qexp:def}).
This is the variational principle.

\section{Legendre transform}

\begin{figure}
\begin {center}
\includegraphics[width=6cm]{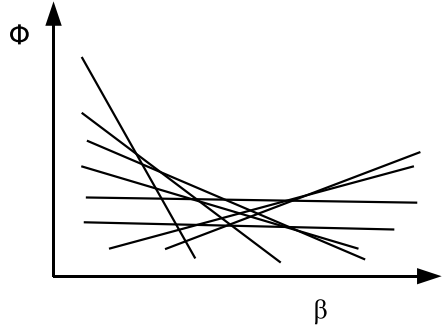}
\end {center}
\caption{$\Phi(\beta)$ determined by taking the supremum of straight lines.}
\label {fig:tangent}
\end{figure}

Note that (\ref {proof:varprin}) is a linear function of $\beta$.
Hence, it determines a straight line in the parameter space.
See the Figure \ref {fig:tangent}.
All these straight lines together determine a convex function
\be
\Phi(\beta)=I(f_\beta)-\beta\int{\rm d}x\,f_\beta(x)H(x).
\label {leg:mass}
\ee
This is Massieu's function.

The thermodynamic entropy $S(U)$ is a function of the internal energy $U$.
Because the latter is a monotonic function of $\beta$ one can make the identifications
\be
S(U)=I_q(f_\beta)\quad\mbox{ and }\quad U=\langle H\rangle_\beta=\int{\rm d}x\,f_\beta(x)H(x).
\ee
Then (\ref {leg:mass}) in combination with (\ref {proof:varprin}) becomes
\be
\Phi(\beta)=S(U)-\beta U=\sup_{U'}\{S(U')-\beta U'\}.
\ee
In particular, this means that Massieu's function $\Phi(\beta)$
is the Legendre transform of the entropy $S(U)$, as is
well known from thermodynamics. An immediate consequence is
\be
\frac {{\rm d}\Phi}{{\rm d}\beta}= -U.
\label {leg:dermass}
\ee
The inverse Legendre transformation is
\be
S(U)=\inf_{\beta'}\{\Phi(\beta')+\beta' U\}.
\ee
This result automatically implies the well-known formula
\be
\frac {{\rm d} S}{{\rm d}U}=\beta.
\label {leg:derent}
\ee

\section {Dual structure}

The equations (\ref {leg:dermass}) and (\ref {leg:derent}) are dual relations in the sense
of thermodynamics. The parameter $\beta$ is the dual of the quantity $U$.
Usually, $\beta$ is the inverse temperature, which is an 'intensive thermodynamic
coordinate', while $U$ is the average total energy and is 'extensive'.
However, the examples of the microcanonical ensemble show that this standard
interpretation is specific for the canonical ensemble.

In a mathematical context the same duality between model parameters and estimators
(averaged quantities used to estimate model parameters) was given a geometric interpretation
by Amari \cite {AS85,MR93,OA07}. His $\alpha$ is related to the deformation index $q$
by $\alpha=1-2q$. The geometric interpretation concerns the statistical manifold,
the definition of which is given in Section \ref {statman}. The flatness of
the statistical manifold is equivalent with the validity of the dual relations
(\ref {leg:dermass}, \ref {leg:derent}).
Many examples found in the literature on nonextensive thermostatistics involve
a curved manifold, which implies that the
parameter $\beta$ does not satisfy (\ref {leg:derent}) and hence, in the context of a canonical
ensemble, does not coincide
with the inverse of the thermodynamic temperature. See for instance \cite {SA06}
and the references quoted there for a discussion about different definitions of temperature
in nonextensive thermostatistics.

Amari's work was the basis for the generalisation found in \cite {NJ04}.
Here, these geometrical insights are reviewed in the context of the
$q$-exponential family.

\section{Estimating inverse temperature}

In principle, knowledge of the average energy $U$ allows determination of the model parameter $\beta$.
A measurement of the total energy $U$ may be very exceptional. However, one can add extra
parameters to the model and measure corresponding quantities to estimate these parameters.
For simplicity only one parameter is considered here.

The value obtained by experimentally measuring $U$ has some uncertainty. It is then obvious to ask
how large is the uncertainty on the estimated parameter $\beta$. This will depend on how large is
the derivative ${\rm d}U/{\rm d}\beta$. Indeed, if $U$ depends only weakly on $\beta$ then
a small error in $U$ leads to a large error in the estimated value of $\beta$.
Now remember that $U$ is minus the derivative of the Massieu function $\Phi$ --- see (\ref {leg:dermass}).
Hence, the relevant quantity is
\be
g(\beta)=\frac {{\rm d}^2\Phi}{{\rm d}\beta^2}.
\label {est:gdef}
\ee
This is called the metric tensor. In the case of multiple parameters it is a matrix.
Because $\Phi(\beta)$ is convex $g(\beta)$ is always positive.

It is known for models belonging to the exponential family that the Fisher information matrix
is equal to the metric tensor. Below it will be shown that this relation can be extended to
models belonging to the $q$-exponential family.

\section{Fisher information}

The $q$-deformed Fisher information is defined by
\be
I_\beta=\int{\rm d}x\,c(x)\left(\frac {c(x)}{f_\beta(x)}\right)^{q}\left(\frac {{\rm d}\,}{{\rm d}\beta}f_\beta(x)\right)^2.
\label {fi:def}
\ee
Note that this definition differs from that studied in \cite {PPM97,PP97,TC98,BPT98,PP04,PP04b}.
It also differs by a scalar factor from the definition given in \cite {NJ04,NJ08},
because in the latter papers the definition is given in terms of a normalised escort probability distribution.
Here, the normalisation is omitted so that the equality $I_\beta=g(\beta)$ holds
without involving a normalisation function.

In order to prove that $I_\beta=g(\beta)$, take the derivative of (\ref {qexpfam:lnexpr}).
This gives
\be
\left(\frac {c(x)}{f_\beta(x)}\right)^{q}\frac {{\rm d}\,}{{\rm d}\beta}\frac {f_\beta(x)}{c(x)}=-\frac {{\rm d}\alpha}{{\rm d}\beta}-H(x).
\label {fi:temp}
\ee
Combining (\ref {fi:temp}) with the definition (\ref {fi:def}) gives
\be
I_\beta&=&\int{\rm d}x\,\left(-\frac {{\rm d}\alpha}{{\rm d}\beta}-H(x)\right)\left(\frac {{\rm d}\,}{{\rm d}\beta}\frac {f_\beta(x)}{c(x)}\right).
\label {fi:temp2}
\ee
But note that $1=\int{\rm d}x\,f_\beta(x)$ implies that
\be
0&=&\frac {{\rm d}\,}{{\rm d}\beta}\int{\rm d}x\,f_\beta(x).
\ee
Hence, (\ref {fi:temp2}) simplifies to
\be
I_\beta&=&-\int{\rm d}x\,H(x)\left(\frac {{\rm d}\,}{{\rm d}\beta}f_\beta(x)\right).
\ee
In combination with (\ref {leg:dermass}) and (\ref {est:gdef}) this yields $I_\beta=g(\beta)$.

\section{The statistical manifold}
\label {statman}

The statistical manifold is now the map
\be
\beta\rightarrow \ln_q\left(\frac {f_\beta(x)}{c(x)}\right).
\ee
It reduces to the log-likelihood function $\beta\rightarrow \ln f_\beta(x)/c(x)$
in the limit $q=1$.
The tangent vector
\be
X_\beta&\equiv&\frac {{\rm d}\,}{{\rm d}\beta}\ln_q\left(\frac {f_\beta(x)}{c(x)}\right)\crcr
&=&-\frac {{\rm d}\alpha}{{\rm d}\beta}-H(x)
\ee
is the generalised score variable. Its average length is defined by
\be
||X_\beta||^2=\int{\rm d}x\,f_\beta(x)^q\left(X_\beta(x)\right)^2.
\label {statman:length}
\ee
A short calculation shows that the latter expression
equals the Fisher information, i.e.~$||X_\beta||^2=I_\beta$.

\section{Final remarks}

In this paper the definition of the generalised exponential family \cite {NJ04,GD04,NJ08}
is considered in the context of nonextensive statistical physics,
where it is called the $q$-exponential family.
Many models of nonextensive statistical physics belong to this family,
while others do not because quite often the normalisation is written
as a prefactor instead of writing it inside the $q$-exponential.
It should be stressed that the prefactor $c(x)$ 
and the Hamiltonian $H(x)$ in the r.h.s.~of (\ref {qexp:def})
must not depend on the parameter $\beta$ while the normalisation
$\alpha(\beta)$ must not depend on the variable $x$.

Several examples of models belonging to the $q$-exponential family
have been given. In particular, the $q$-Gaussians can be written
in the required form (\ref {qexp:def}). The $q$-Gaussian model
receives a lot of interest because it appears as the central limit
of strongly correlated models --- see for instance \cite {MTG06,HS07,UT07,VP07,VP07b,TH08}.
To my knowledge, the examples concerning the microcanonical ensemble
appear in the literature for the first time.

The role of the escort probabilities \cite {TMP98} has not been discussed. But the
unnormalised escort probabilities $f_\beta(x)^q$ appear prominently, for instance in (\ref {statman:length}).
By leaving out the normalisation in the definition (\ref {fi:def}) of the Fisher information
the metric tensor $g$ equals the Fisher information, while in \cite {NJ04,NJ08}
the normalisation factor enters as a multiplicative factor.

Only continuous distributions have been considered here. The translation
towards discrete probability distributions is straigthforward.
The transition to quantum models requires more attention but is feasible.
An early step in this direction is found in \cite {NJ05}.
In particular, the quantum analogue of (\ref {choice:genent}) is
$I(\rho)=-\Tr F(\rho)$. The prior weights $c(x)$ must be taken
all equal before making the transition to quantum mechanics because
cyclic permutation under the trace is essential --- see \cite {NJu}.

The presentation has been restricted to single parameter models.
The extension to more than one parameter is obvious. Note that
in the mathematics literature also non-parametric models
are considered \cite {HH93,PS95}.

Some topics have been left out of the paper.
In particular, the relative entropy of the Bregman type was not mentioned.
Neither was the relation between Fisher information and the inequality
of Cram\'er and Rao. Both can be found in \cite {NJ04} in the more
general context. Finally, note that one can expect
that the generalisations discussed in the present paper,
and, in particular, the geometric insight behind them,
may lead to powerful applications.

\section*{Appendix}

For convenience, explicite expressions used in the examples of Section 3
have been brought together in the following table

\begin{center}
\begin{tabular}{rcccc}
$q$           & \vline & $\ln_q(u)$ & $\exp_q(u)$ & $F_q(u)$\\
\hline
 1            & \vline & $\ln u$                               & $e^u$                            & $u\ln u$\\
 2            & \vline & $1-\frac 1u$                          & $\frac 1{[1-y]_+}$               & $u-\ln u+A-1$ \\
 3            & \vline & $\frac 12\left(1-\frac 1{u^2}\right)$ & $\frac 1{[1-2u]_+^{1/2}}$        & $\frac 12\left(u+\frac 1u\right) +A-1$\\
 $\frac 12$   & \vline & $2(\sqrt u-1)$                        & $\left[1+\frac 12u^2\right]_+^2$ & $2u\left(\frac 23\sqrt u-1\right)$  \\
 $-1$         & \vline & $\frac 12(u^2-1)$                     & $[1+2u]_+^{1/2}$                 & $\frac 12 u\left(\frac 13u^2-1\right)$
\end{tabular}
\end{center}

The corresponding expressions for the entropy functional $I_q(f)$ are
\be
I_1(f)&=&-\int{\rm d}x\,f(x)\ln\frac {f(x)}{c(x)},\\
I_2(f)&=&-1+\int{\rm d}x\,c(x)\left(\ln \frac {f(x)}{c(x)}-A+1\right) \\
I_3(f)&=&-\frac 12-\int{\rm d}x\,c(x)\left(\frac {c(x)}{f(x)}+A-1\right),\\
I_{1/2}(f)&=&-2\int{\rm d}x\,f(x)\left(\sqrt{\frac {f(x)}{c(x)}}-1\right),\\
I_{-1}(f)&=&-\frac 12\int{\rm d}x\,f(x)\left(\frac 13\left(\frac {f(x)}{c(x)}\right)^2-1\right).
\ee

\section*{}

\end{document}